\documentclass[a4paper,11pt]{article}
\usepackage{amsmath}
\usepackage{amssymb}
\usepackage{latexsym}
\usepackage{enumerate}
\usepackage{graphicx}
\usepackage{placeins}
\usepackage{multirow}
\usepackage{authblk}
\usepackage{blindtext}

\begin{document}

\date{\today }
\title{On the energy of a non-singular black hole solution satisfying the weak energy condition}

\author[1]{I. Radinschi\thanks{radinschi@yahoo.com}}
\affil[1]{Department of Physics ,
 ``Gheorghe Asachi''   Technical University,
Iasi, 700050, Romania}

\author[2]{Th. Grammenos\thanks{thgramme@civ.uth.gr}}
\affil[2]{Department of Civil Engineering,
University of Thessaly, 383 34 Volos, Greece}

\author[3]{F. Rahaman\thanks{rahaman@associates.iucaa.in}}
\affil[3]{Department of
Mathematics, Jadavpur University, Kolkata 700 032, West Bengal,
India.}

\author[1]{M. M. Cazacu\thanks{marius.cazacu@tuiasi.ro}}

\author[4]{A. Spanou\thanks{aspanou@central.ntua.gr}}
\affil[4]{School of Applied Mathematics and Physical Sciences,
National Technical University of Athens, 157 80 Athens, Greece}

\author[3]{J. Chakraborty\thanks{joydeep\_sweety2010@rediffmail.com}}

\maketitle

\begin{abstract}
The energy-momentum localization for a new four-dimensional and spherically symmetric, charged black hole solution that through a coupling of general relativity with non-linear electrodynamics is everywhere non-singular while it satisfies the weak energy condition is investigated. The Einstein and M\o ller
energy-momentum complexes have been employed in order to calculate the energy distribution and the momenta for the aforesaid solution. It is found that the energy distribution depends explicitly on the mass and the charge of the black hole, on two parameters arising from the space-time geometry considered, and on the radial coordinate. Further, in both prescriptions all the momenta vanish.In addition, a comparison of the results obtained by the two energy-momentum complexes is made, whereby some limiting and particular cases are pointed out.
\end{abstract}

\section{Introduction}

Even though the problem of energy-momentum localization has triggered a lot of interesting research work, it still remains not fully answered for more than a century. The only step one can hope to make forward is to find a more powerful tool for accessing this issue of general relativity.

The use of different tools for energy-momentum localization, like
super-energy tensors \cite{Bel}-\cite{Senovilla}, quasi-local definitions \cite{Brown}-\cite{Balart_1} and the
energy-momentum complexes \cite{Einstein}-\cite{Weinberg} has led to the development of several
interesting works. In particular, the energy-momentum complexes of Einstein \cite{Einstein},
Landau-Lifshitz \cite{Landau}, Papapetrou \cite{Papapetrou}, Bergmann-Thomson \cite{Bergmann} and Weinberg \cite{Weinberg} are pseudotensorial quantities and coordinate dependent. They can be used in Cartesian and quasi-Cartesian coordinates, more precisely
in Schwarzschild Cartesian coordinates and in Kerr-Schild Cartesian
coordinates, and have yielded so far many physically meaningful results \cite{Virbhadra1990a}-\cite{Tripathy}. 
The M\o ller energy-momentum complex \cite{Moller} allows the calculation of energy and momenta in any coordinate system, including Schwarzschild
Cartesian coordinates and Kerr-Schild Cartesian coordinates, and has also
provided physically interesting results  for many
space-time geometries, in  particularly for ($3+1$), ($2+1$)
and ($1+1$) space-times \cite{Ching-Ra}-\cite{Mat}.
It is worth noting that different pseudotensors
yield the same energy value for any metric of the Kerr-Schild class and
also for solutions more general than those of the Kerr-Schild class (for
reviews and references, see the works \cite{Aguirregabiria}, \cite{Xulu_2} and \cite{Virb1999}).
Moreover, the localization of energy-momentum has been studied 
in the context of teleparallel theory of gravitation whereby many similar results have been obtained \cite{Moller1964}-\cite{Ganiou}.

The Einstein, Landau-Lifshitz, Papapetrou, Bergmann-Thomson, Weinberg and M\o ller prescriptions are in agreement with the definition of the
quasi-local mass given by Penrose \cite{Penrose} and developed by Tod \cite{Tod} for some
gravitating systems.
Recently, the new concept in the localization of energy is that of
quasi-local energy-momentum associated with a closed 2-surface. In fact, it has
been demonstrated that ``the quasi-local quantities could provide a more
detailed characterization of the states of the gravitational field" \cite{Szabados}.
The energy localization is also
connected with the quasi-local energy given by Wang and Yau \cite{Wang2009a}, \cite{Wang2009b}. The
rehabilitation of energy-momentum complexes concerns the searching for a
common quasi-local energy value. Recently, an important discovery has been
made, namely,  by considering pseudotensors and quasi-local approaches in the context of the Hamiltonian formulation and with the choice of a 4D isometric Minkowski reference geometry on the boundary, it is found that for any closed 2-surface there is a common value for the quasi-local energy for all expressions that are in agreement (to linear order) with the Freud superpotential or, put simply, all the quasi-local expressions in a large class yield the same energy-momentum \cite{Nester2018a}, \cite{Nester2018b}.

This paper is organized as follows. In Section 2 we briefly present the non-singular black hole solution that satisfies the weak energy condition.
In Section 3 the two prescriptions of Einstein and M\o %
ller used for the calculation of the energy distribution and
momenta are introduced. In Section 4 we show the results obtained for the aforementioned non-singular black hole solution by applying the pseudotensorial
definitions given in Section 3. Finally, in the Discussion we summarize the
results and we focus on commenting about some limiting and particular cases. Throughout we have used geometrized units ($c=G=1$) and for the signature the
choice has been ($+$,$-$,$-$,$-$). In the case of the Einstein prescription
we have used the Schwarzschild Cartesian coordinates $\{t$, $x$, $y$, $z\}$
and for the M\o ller prescription the Schwarzschild coordinates $\{t,$ $r,$ $%
\theta ,$ $\phi \}$, respectively. Further, Greek indices take values from $0$
to $3$, while Latin indices run from $1$ to $3$.

\section{The Non-Sinular Black Hole Solution Satisfying the Weak Energy Condition}

In this section the new spherically symmetric and charged non-singular black
hole solution that satisfies the weak energy condition developed by L. Balart and E. C.
Vagenas \cite{Vagenas2014} is introduced. This black hole solution has been found in the context of general relativity coupled to non linear electrodynamics via a term $L(F)$ in the action, with the Lorentz-invariant scalar $F=\frac{1}{4}F^{\mu\nu}F_{\mu\nu}$, and $F_{\mu\nu}$ is the electromagnetic tensor that reduces to the Maxwellian field tensor in the weak field case. Here, $F_{\mu\nu}$ is restricted only to the electric field. The black hole metric has been constructed by employing the
Dagum distribution \cite{Dagum}, a continuous probability distribution, given by
\begin{equation}
\sigma (x)=\frac{ap\,x^{ap-1}}{b^{ap}\left[1+(\frac{x}{b})^{a}\right]^{p+1}}, 
\tag{1}
\end{equation}
where $x>0$ , while the parameters $a$, $b$, $p$ $\in \mathbb{R}^+$. For $p=\frac{1}{a}$ and $b=1$, $\sigma(x)$ can be written as
\begin{equation}
\sigma (x)=\frac{1}{(1+x^{a})^{\frac{a+1}{a}}}.  \tag{2}
\end{equation}
Here, with the transformation $x\rightarrow \frac{q^{2}}{Mr}$  the black hole metric function reads
\begin{equation}
f(r)=1-\frac{2M}{r}\left\{\frac{1}{\left[1+\gamma \left( \frac{q^{2}}{Mr}%
\right)^{a}\right]^{\frac{a+1}{a}}}\right\}^{\beta },  \tag{3}
\end{equation}
with $\gamma\in \mathbb{R}^+$, while $M$ and $q$ are the mass and the charge of the black hole, respectively. For $\beta \geq \frac{3}{(a+1)}$ this black hole metric is
non-singular everywhere, while for $\beta \leq \frac{3}{(a+1)}$ it satisfies
the weak energy condition $T_{\mu\nu}u^{\mu}u^{\nu}\geq 0$ for every timelike vector $u^{\mu}$, according to which the local energy density must be positive-definite. This general inequality is shown in \cite{Vagenas2014b} to be equivalent to two inequalities on the first and second derivative of the mass function w.r.t. $r$. In order to have a non-singular black hole solution that also satisfies the weak energy condition, one imposes the condition  $\beta =\frac{3}{(a+1)}$, so that the obtained new
spherically symmetric, static and charged non-singular  black hole metric has a line element of the form 
\begin{equation}
ds^{2}=B(r)dt^{2}-A(r)dr^{2}-r^{2}(d\theta ^{2}+\sin ^{2}\theta d\phi ^{2}),
\tag{4}
\end{equation}
with $B(r)=f(r)$, $A(r)=\frac{1}{f(r)}$,  and the metric function now reads
\begin{equation}
f(r)=1-\frac{2M}{r}\left[ \frac{1}{1+\gamma \left( \frac{q^{2}}{Mr}%
\right)^{a}}\right]^{3/a}.  \tag{5}
\end{equation}

The corresponding electric field is
\begin{equation}
E(r)=\frac{3q}{2r^{2}}\frac{\gamma (3+a)(\frac{q^{2}}{Mr})^{a-1}}{%
\left[1+\gamma \left( \frac{q^{2}}{Mr}\right)^{a}\right]^{2+3/a}}.  \tag{6}
\end{equation}

For small values of the radial coordinate $r$ the family of metrics derived from (5)
behaves as a de Sitter black hole
\begin{equation}
f(r)\approx 1-\frac{2\,M^{4}}{\gamma ^{3/a}\,q^{6}}r^{2},  \tag{7}
\end{equation}
while asymptotically it behaves as the Schwarzschild black hole.

It is important to notice that a particuar case is obtained for $a=1$ (hence $\beta=3/2$) and $%
\gamma =\frac{1}{6}$, so that the metric function (5) reads
\begin{equation}
f(r)=1-\frac{2\,M}{r}\left( \frac{1}{1+\frac{q^{2}}{6M\,r}}\right) ^{3}
\tag{8}
\end{equation}
and the black hole solution asymptotically behaves as the Reissner-Nordstr\"{o}m solution.
For all the other values of the parameters $a$ and $\gamma$ the non-singular
black hole space-time geometries with the metric function (5) asymptotically do not have the behaviour of the
Reissner-Nordstr\"{o}m solution. It must also be pointed out that only for the values $a=1$ and $\gamma =\frac{1}{6}$ one gets a nonlinear electrodynamics model leading to Maxwell's electrodynamics in the weak field
approximation.

\section{Einstein and\ M\o ller Prescriptions}

The Einstein energy-momentum complex \cite{Einstein} defined for a ($3+1$) dimensional
space-time is given by 
\begin{equation}
\theta _{\nu }^{\mu }=\frac{1}{16\pi }h_{\nu ,\,\lambda }^{\mu \lambda }. 
\tag{9}
\end{equation}%
The superpotentials $h_{\nu }^{\mu \lambda }$ are expressed as 
\begin{equation}
h_{\nu }^{\mu \lambda }=\frac{1}{\sqrt{-g}}g_{\nu \sigma }\left[ -g(g^{\mu
\sigma }g^{\lambda \kappa }-g^{\lambda \sigma }g^{\mu \kappa })\right]
_{,\kappa }  \tag{10}
\end{equation}%
and satisfy the necessary antisymmetric property 
\begin{equation}
h_{\nu }^{\mu \lambda }=-h_{\nu }^{\lambda \mu }.  \tag{11}
\end{equation}%
In the Einstein prescription the local conservation law holds: 
\begin{equation}
\theta _{\nu ,\,\mu }^{\mu }=0.  \tag{12}
\end{equation}%
The energy and momentum can be calculated with 
\begin{equation}
P_{\mu }=\iiint \theta _{\mu }^{0}\,dx^{1}dx^{2}dx^{3},  \tag{13}
\end{equation}%
where $\theta _{0}^{0}$ and $\theta _{i}^{0}$ represent the energy and
momentum density components, respectively.

Applying Gauss' theorem, the energy-momentum becomes
\begin{equation}
P_{\mu }=\frac{1}{16\pi }\iint h_{\mu }^{0i}n_{i}dS,  \tag{14}
\end{equation}
with $n_{i}$ the outward unit normal vector over the surface $dS.$ In eq.
(14) $P_{0}$ represents the energy.

The expression for the M{\o }ller energy-momentum complex \cite{Moller} is given by

\begin{equation}
\mathcal{J}_{\nu }^{\mu }=\frac{1}{8\pi }M_{\nu \,\,,\,\lambda }^{\mu
\lambda },  \tag{15}
\end{equation}%
with the M{\o }ller superpotentials $M_{\nu }^{\mu \lambda }$ 
\begin{equation}
M_{\nu }^{\mu \lambda }=\sqrt{-g}\left( \frac{\partial g_{\nu \sigma }}{%
\partial x^{\kappa }}-\frac{\partial g_{\nu \kappa }}{\partial x^{\sigma }}%
\right) g^{\mu \kappa }g^{\lambda \sigma }.  \tag{16}
\end{equation}%
The M{\o }ller superpotentials $M_{\nu }^{\mu \lambda }$ are also
antisymmetric
\begin{equation}
M_{\nu }^{\mu \lambda }=-M_{\nu }^{\lambda \mu }.  \tag{17}
\end{equation}%
M{\o }ller's energy-momentum complex also satifies the local conservation law%
\begin{equation}
\frac{\partial \mathcal{J}_{\nu }^{\mu }}{\partial x^{\mu }}=0.  \tag{18}
\end{equation}%
In (15) $\mathcal{J}_{0}^{0}$ is the energy density and $\mathcal{J}_{i}^{0}$
represents the momentum density components.

For the M\o ller prescription, the energy and momentum distributions are
given by 
\begin{equation}
P_{\mu }=\iiint \mathcal{J}_{\mu }^{0}dx^{1}dx^{2}dx^{3}  \tag{19}
\end{equation}%
and the energy distribution can be evaluated with 
\begin{equation}
E=\iiint \mathcal{J}_{0}^{0}dx^{1}dx^{2}dx^{3}.  \tag{20}
\end{equation}%
Using Gauss' theorem one obtains
\begin{equation}
P_{\mu }=\frac{1}{8\pi }\iint M_{\mu }^{0i}n_{i}dS.  \tag{21}
\end{equation}

\section{Energy-Momentum Distribution of the Non-Singular Black Hole Solution
Satisfying the Weak Energy Condition}

In order to perform the calculations using
the Einstein prescription, the metric given by the line element (4) is transformed 
into Schwarzschild Cartesian coordinates applying the coordinate
transformation $x=r\,\sin \theta \cos \varphi ,$ $y=r\,\sin \theta \sin
\varphi ,$ $z=r\,\cos \theta $. Then, the line element takes the following form:
\begin{equation}
ds^{2}=B(r)dt^{2}-(dx^{2}+dy^{2}+dz^{2})-\frac{A(r)-1}{r^{2}}%
(xdx+ydy+zdz)^{2}\text{.}  \tag{22}
\end{equation}

The calculations for $\mu = 0, 1,2,3$ and 
$i=1,2,3$ for the components of the superpotential $h_{\mu }^{0i}$ (needed in (14)) in
quasi-Cartesian coordinates are
\begin{equation} \tag{23}
\begin{split}
& h_{1}^{01}=h_{1}^{02}=h_{1}^{03}=0, \\
& h_{2}^{01}=h_{2}^{02}=h_{2}^{03}=0,\\
& h_{3}^{01}=h_{3}^{02}=h_{3}^{03}=0
\end{split}
\end{equation}
while the non-vanishing components of the superpotential are
\begin{equation}
h_{0}^{01}=\frac{4Mx}{r^{3}}\left[\frac{1}{1+\gamma \left( 
\frac{q^{2}}{Mr}\right) ^{a}}\right]^{3/a},  \tag{24}
\end{equation}

\begin{equation}
h_{0}^{02}=\frac{4My}{r^{3}}\left[ \frac{1}{1+\gamma \left( 
\frac{q^{2}}{Mr}\right) ^{a}}\right]^{3/a},  \tag{25}
\end{equation}

\begin{equation}
h_{0}^{03}=\frac{4Mz}{r^{3}}\left[ \frac{1}{1+\gamma \left( 
\frac{q^{2}}{Mr}\right) ^{a}}\right]^{3/a}.  \tag{26}
\end{equation}

With the aid of the line element (22), the expression for the energy-momentum
distribution (14) and the expressions (24)-(26) for the superpotentials, the
energy distribution in the Einstein prescription for the new charged non-singular black hole solution satisfying the weak energy condition is obtained as
\begin{equation}
E_{E}=M\left[ \frac{1}{1+\gamma \left( \frac{q^{2}}{Mr}\right) ^{a}}%
\right] ^{3/a},  \tag{27}
\end{equation}
while, due to (23), we find that all the momentum components vanish:
\begin{equation}
P_{x}=P_{y}=P_{z}=0.  \tag{28}
\end{equation}

\begin{figure}[h!]
\centering
\includegraphics[width=7cm]{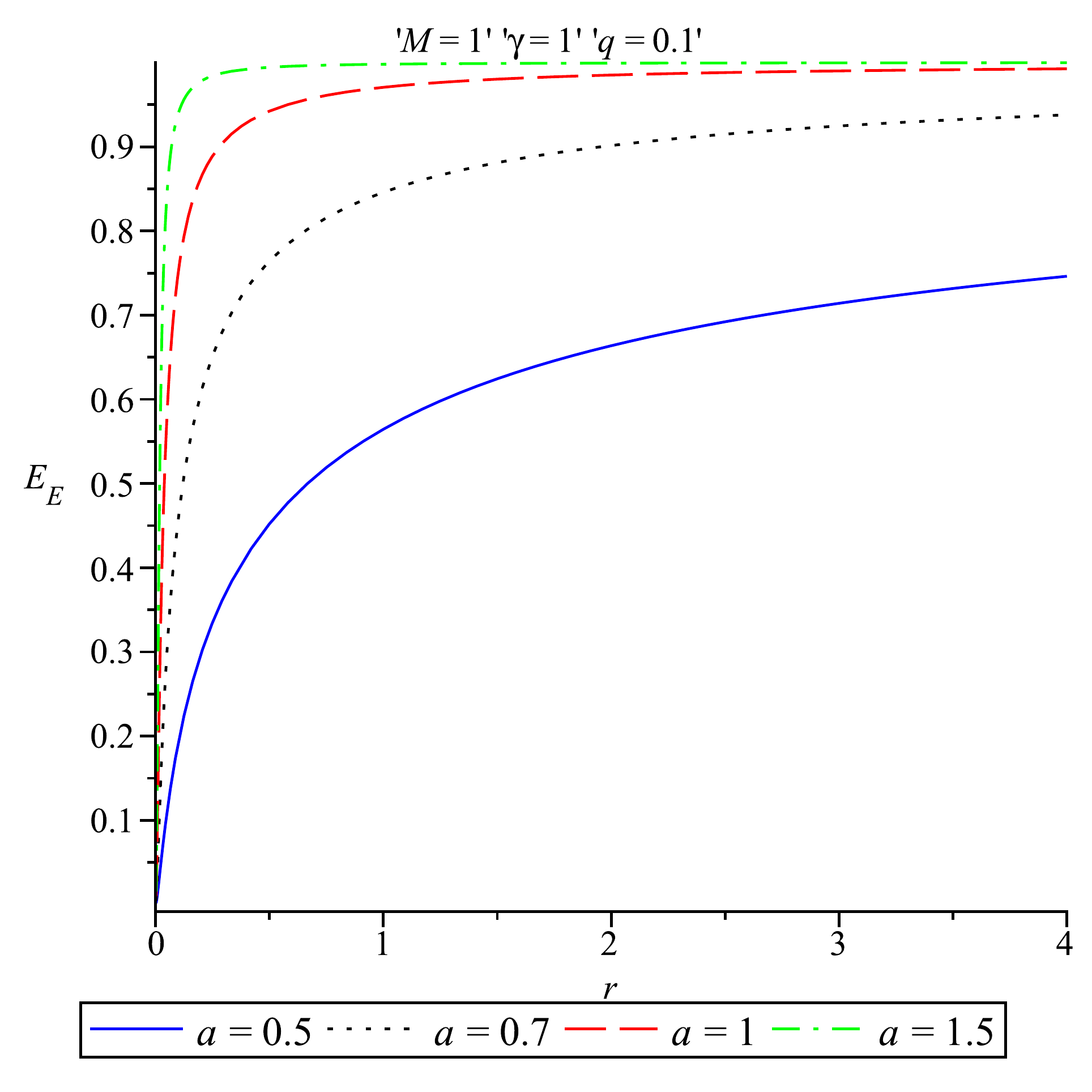}
\caption{Einstein energy vs.\ $r$ for various values of the parameter $a$ and $\gamma=1$.}
\label{fig1}
\end{figure}

In Fig. 1, the energy distribution (27) obtained  in the Einstein prescription is plotted as a function of $r$ for
four different values of the parameter $a$ and $M=1$, $q=0.1$, and $\gamma =1$.

Fig. 2 exhibits the behaviour of the Einstein energy distribution as a function of $r$ near the origin
for four different values of $a$ and $M=1$, $q=0.1$, and $\gamma =1$.

\begin{figure}[h!]
\centering
\includegraphics[width=7cm]{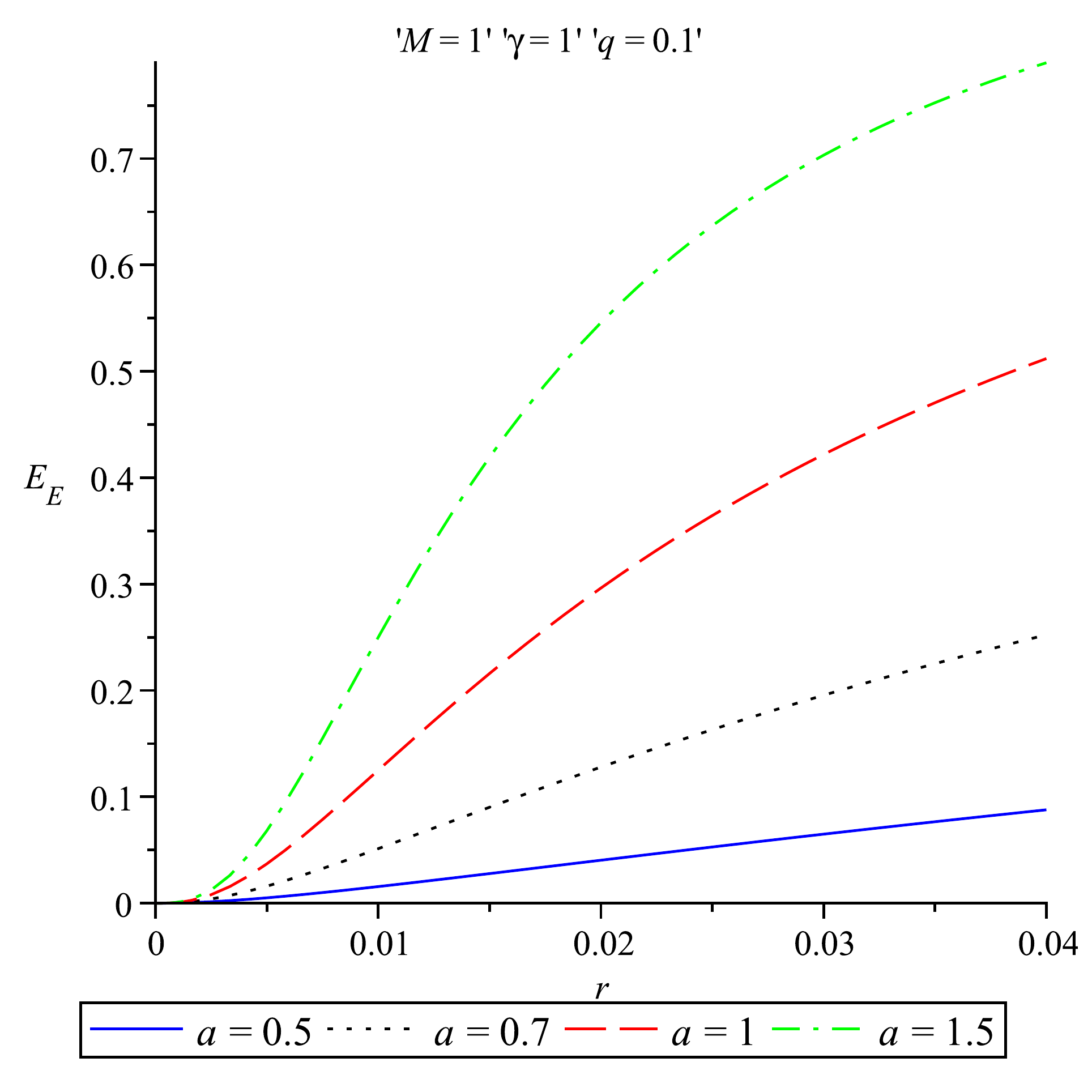}
\caption{Einstein energy vs.\ $r$ near the origin for various values of the parameter $a$ and $\gamma=1$.}
\label{fig2}
\end{figure}

Next, we apply the M\o ller prescription using  Schwarzschild
coordinates $\{t,$ $r,$ $\theta ,$ $\phi \}$, for the line element (4)
and the metric function (5).  The only non-vanishing component of the M\o ller superpotential  (16) is found to be
\begin{equation}\tag{29}
M_{0}^{01}=M\left[\frac{1}{1+\gamma \left(\frac{q^2}{Mr}\right)^a}\right]^{3/a}
\left[
2-\frac{6\gamma\left(\frac{q^2}{Mr}\right)^a}{1+\gamma\left(\frac{q^2}{Mr}\right)^a}
\right]\sin\theta
\end{equation}
while all the other components vanish.

Inserting the above expression (29) of the M\o ller superpotential into (21), we obtain 
the energy distribution in the M\o ller prescription:
\begin{equation}\tag{30}
\begin{split}
E_{M} &= M\left[\frac{1}{1+\gamma \left(\frac{q^2}{Mr}\right)^a}\right]^{3/a}
\left[
1-\frac{
3\gamma\left(\frac{q^2}{Mr}\right)^a}{1+\gamma \left( 
\frac{q^{2}}{Mr}\right) ^{a}}\right]
\\
& =
E_E\left[
1-\frac{
3\gamma\left(\frac{q^2}{Mr}\right)^a}{1+\gamma \left( 
\frac{q^{2}}{Mr}\right) ^{a} }
\right],
\end{split}
\end{equation}
from which we infer that the energy in the M\o ller prescription can be given in terms of the energy in the Einstein prescription.

\begin{figure}[!t]
\centering
\includegraphics[width=7cm]{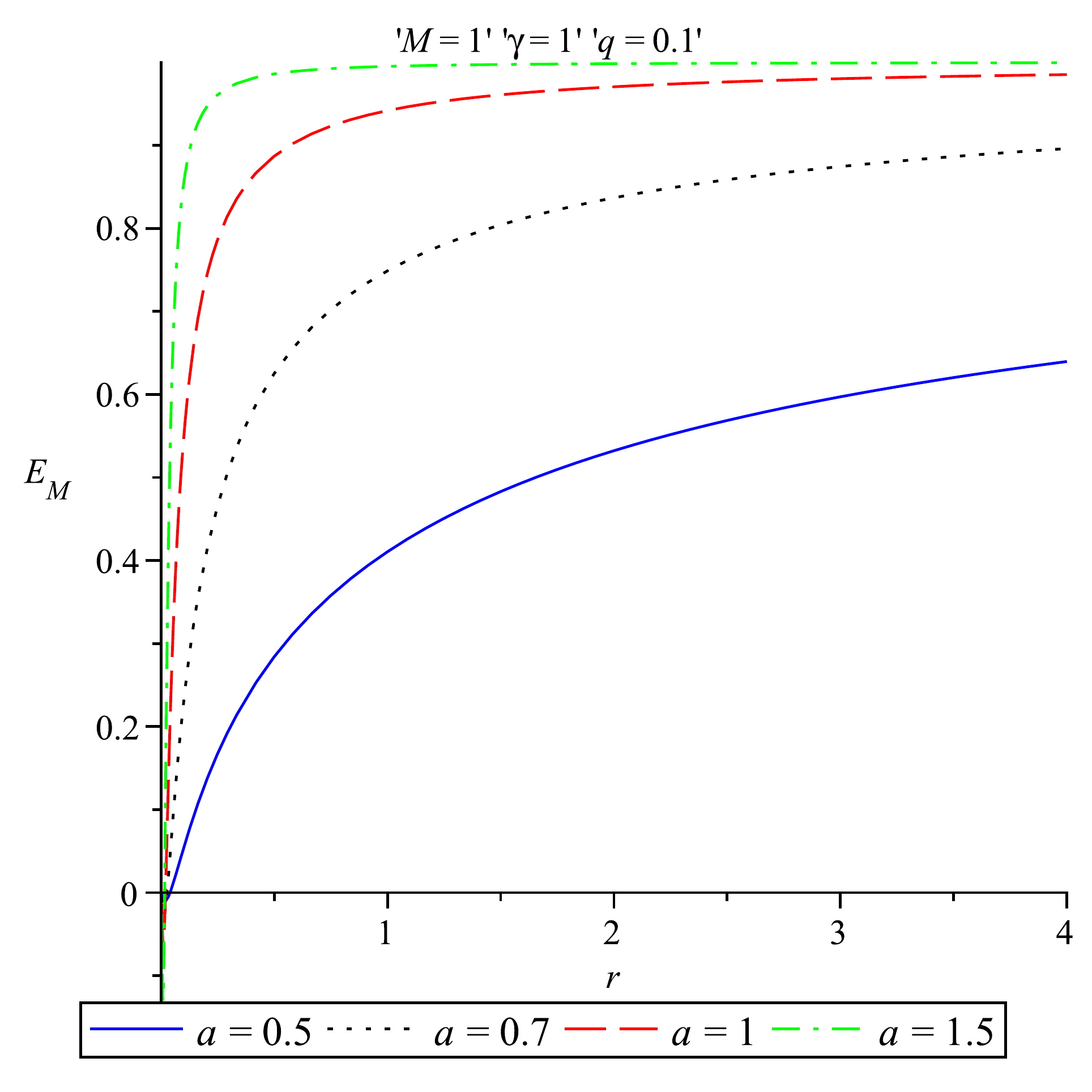}
\caption{M\o ller energy vs.\ $r$ for various values of the parameter $a$ and $\gamma=1$.}
\label{fig1}
\end{figure}
\begin{figure}[!t]
\centering
\includegraphics[width=7cm]{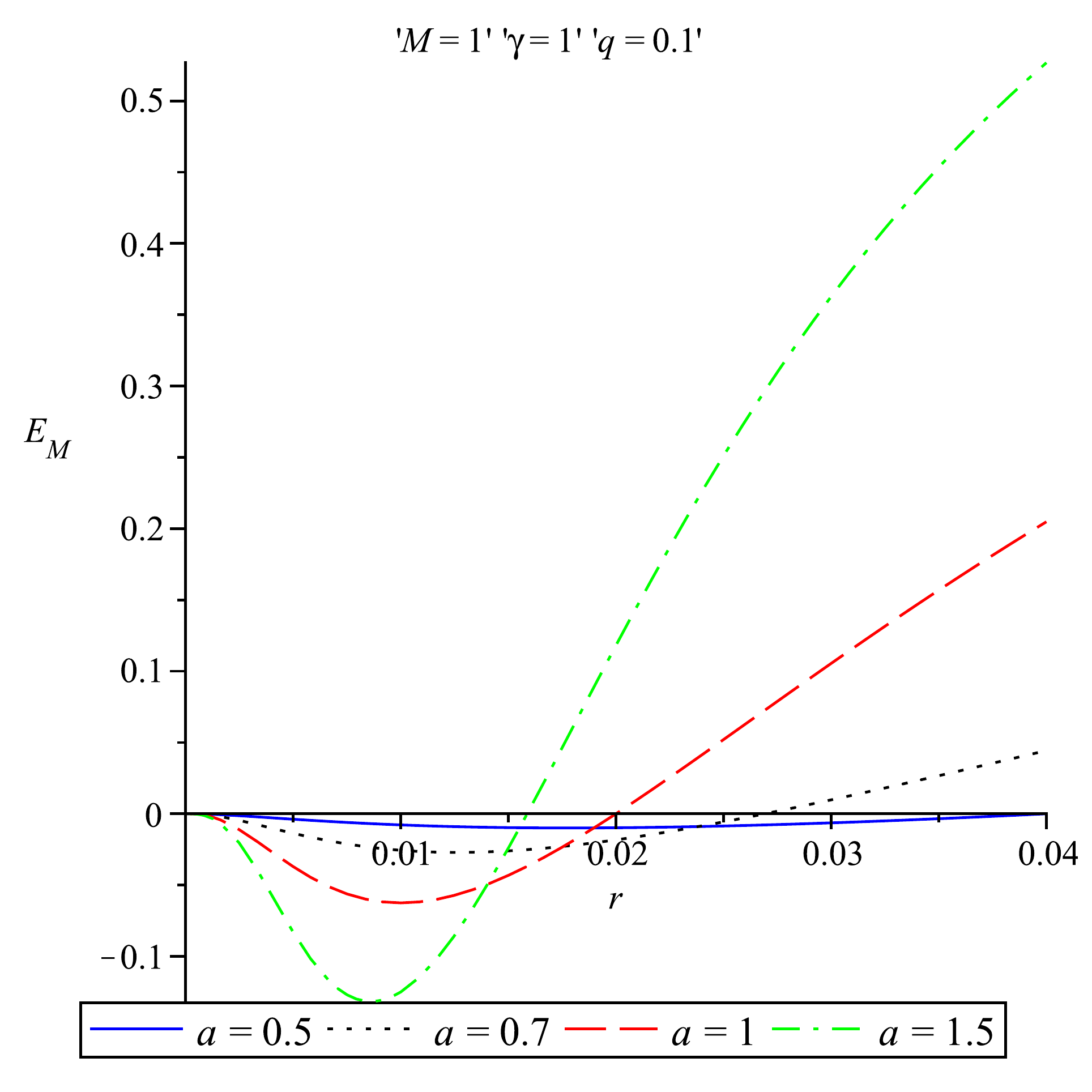}
\caption{M\o ller energy vs.\ $r$ near the origin for various values of the parameter $a$ and $\gamma=1$.}
\label{fig1}
\end{figure}

Furthermore, as expected from the vanishing of the spatial components of the M\o ller superpotential, all the momentum components are found to be zero:
\begin{equation}
P_{r}=P_{\theta }=P_{\phi }=0.  \tag{31}
\end{equation}

\begin{figure}[!h]
\centering
\includegraphics[width=7cm]{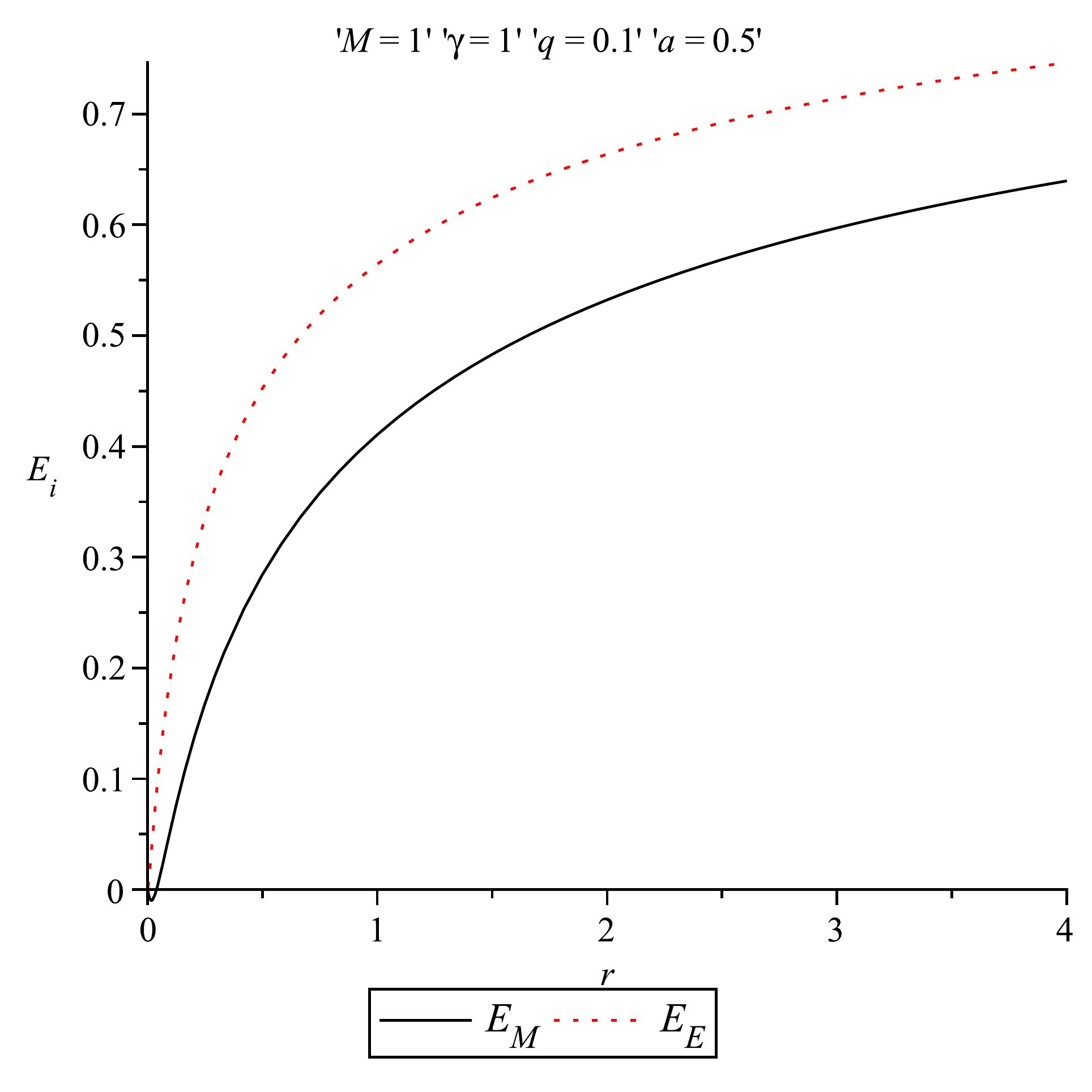}
\caption{Energy distributions in the
Einstein and M\o ller prescriptions for $\gamma=1$ and $a=0.5$.}
\label{fig1}
\end{figure}

\begin{figure}[!h]
\centering
\includegraphics[width=7cm]{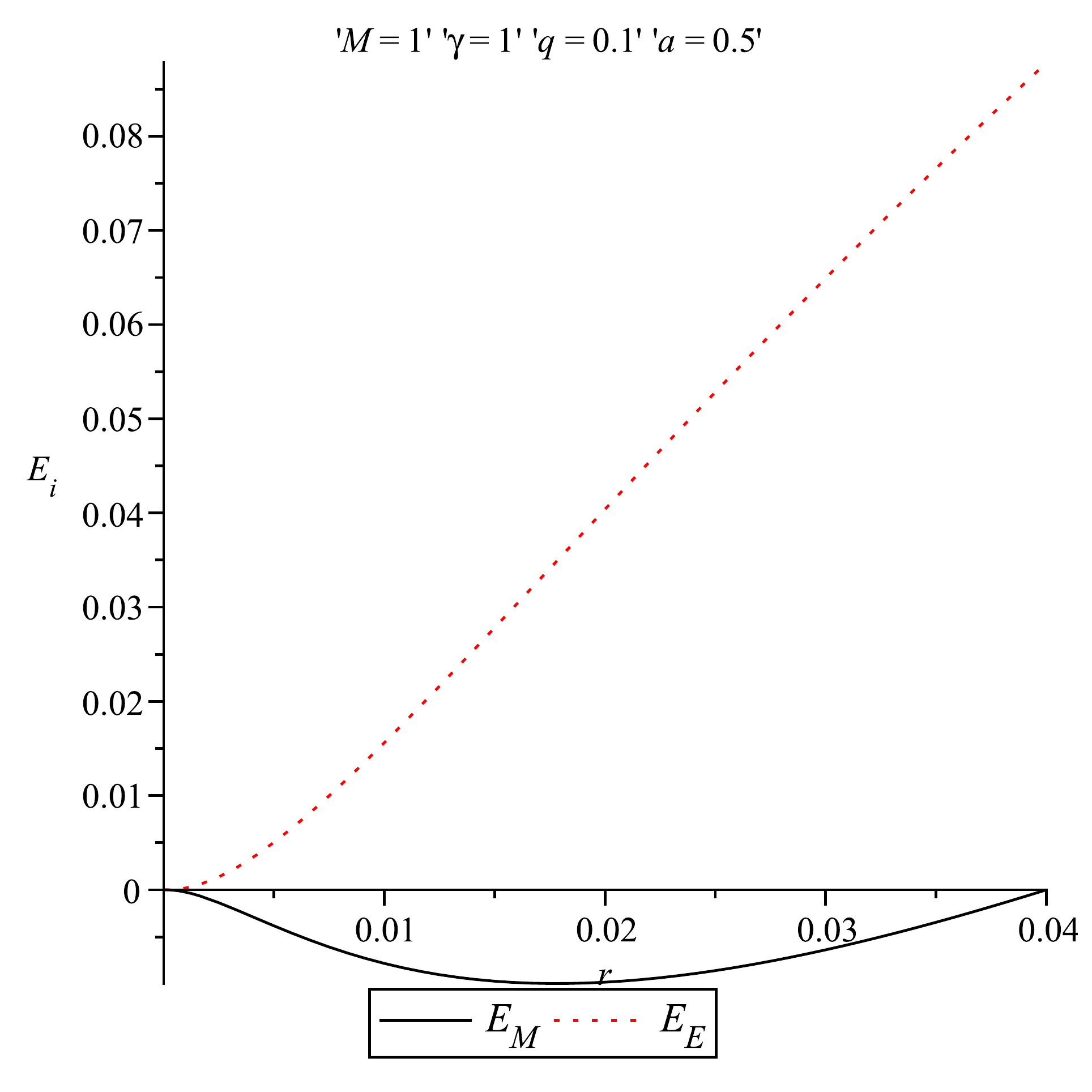}
\caption{Energy distributions in the
Einstein and M\o ller prescriptions for $\gamma=1$ and $a=0.5$ near origin.}
\label{fig1}
\end{figure}

In Fig. 3 the energy distribution in the M\o ller prescription for
four different values of the parameter $a$ and $M=1$, $q=0.1$, and $\gamma =1$ is plotted, while in Fig.~4 
the behaviour of the M\o ller energy distribution near the origin is presented 
for four different values of the parameter $a$ and $M=1$, $q=0.1$, and $\gamma =1$.

A comparison of the energy distributions in the Einstein and M\o ller prescriptions is given in Fig.~5 for $a=0.5$ and  $M=1$, $q=0.1$, and $\gamma =1$. 
Additionally, in Fig. 6 an analogous comparison of the energy distributions in
the Einstein and M\o ller prescriptions near the origin is presented for the same values of the parameters $a$,  $M$, $q$,  and $\gamma$.

\section{Discussion}

The aim of this work is the study of the energy-momentum distribution for
a new spherically symmetric and charged, non-singular black hole solution satisfying the weak energy condition.

To this purpose, the Einstein and M\o ller energy-momentum complexes
have been applied and it has been found that all the momenta vanish in both pseudotensorial prescriptions. Additionally, the energy distributions obtained have well-defined expressions 
showing a dependence on the mass $M$, the charge $q,$ the two parameters $\gamma $ and $a$, introduced in Section 2, and on the radial coordinate $r$. Furthermore, the M\o ller energy is expressed in terms of the Einstein energy. Both energies acquire the same value $M$ (ADM mass) for $r\rightarrow\infty$ or for $q=0$. 

We have also studied the limiting behavior of the energy distribution for $%
r\rightarrow 0$ and $r\rightarrow \infty $, and for the particular cases $%
q=0 $ and $a=1$ and $\gamma =\frac{1}{6}$, respectively. In Table 1 we
summarize the physically meaningful results for these limiting and
particular cases.%
\begin{table}
\centering
\begin{tabular}{|c|c|c|c|c|}
\hline
Case & $r\rightarrow 0$ & $r\rightarrow \infty $ & $q=0$ & $a=1$, $\gamma =%
\frac{1}{6}$ \\ 
\hline
&&&&\\
$E_{E}$ & $0$ & $M$ & $M$ & $M\left( \frac{1}{1+\frac{q^{2}}{6Mr}}\right)
^{3}$ \\ 
&&&&\\

$E_{M}$ & $0$ & $M$ & $M$ & $E_{E}\left[1-\frac{\frac{q^{2}}{Mr}}{2\left( 1+\frac{q^{2}}{6Mr}\right) }\right]$
\\
&&&&\\
 \hline
\end{tabular}
\caption{Limiting and particular values for the Einstein energy $E_E$ and the M\o ller energy $E_M$.}
\end{table}

At this point some remarks are need to clarify the results. From Table 1 we conclude that the energy distribution in both prescriptions vanishes near the
origin $r\rightarrow 0$, while for either $r\rightarrow \infty $ or $q=0$ both energies are equal to the ADM mass $M$, in agreement to the result obtained by Virbhadra for the energy
distribution of the Schwarzschild black hole solution \cite{Virb1999}. 
For the particular case $a=1$ and $\gamma =\frac{1}{6}$ we get an expression for
the energy distribution that depends on the mass $M$ and the charge $q$ of the black hole,
as well as on  the radial coordinate $r$ for a model of general relativity coupled to 
nonlinear electrodynamics, the latter corresponding to  Maxwell's theory in the weak field approximation. 
Furthermore, the non-singular black hole solution with $%
\beta =\frac{3}{2}$, $a=1$ and $\gamma =\frac{1}{6}$ satisfies the weak energy condition and
is described by the metric function $f(r)=1-\frac{2\,M}{r}\left( \frac{1}{1+%
\frac{q^{2}}{6M\,r}}\right) ^{3}$, while asymptotically it behaves as the
Reissner-Nordstr\"{o}m solution. In fact, for $r\rightarrow 0$ this non-singular and charged black hole solution
exhibits a de Sitter behaviour near zero.

\begin{figure}[h!]
\centering
\includegraphics[width=7cm]{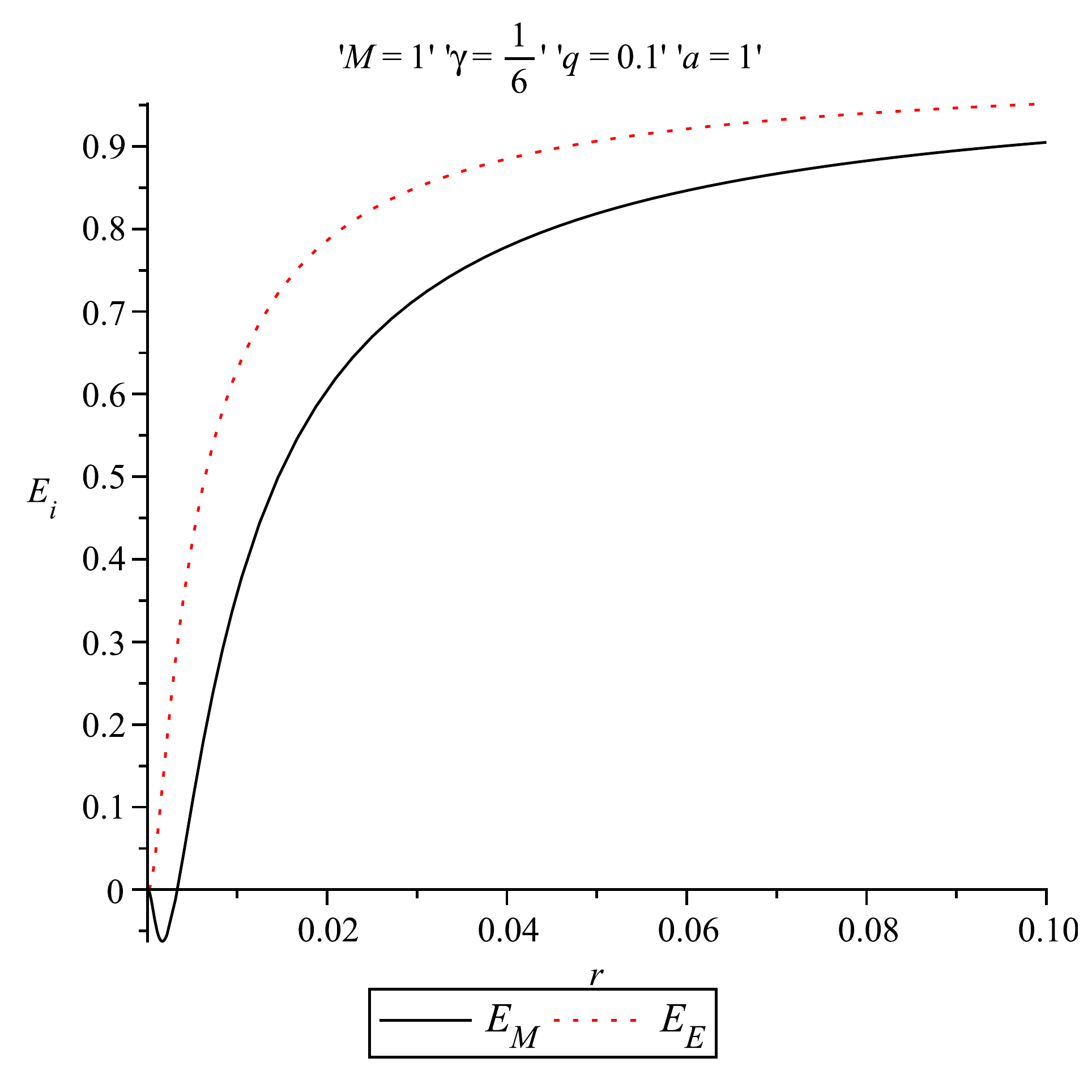}
\caption{Energy distributions in the
Einstein and M\o ller prescriptions for $a=1$ and $\gamma =\frac{1}{6}$.}
\label{fig1}
\end{figure}

\begin{figure}[h!]
\centering
\includegraphics[width=7cm]{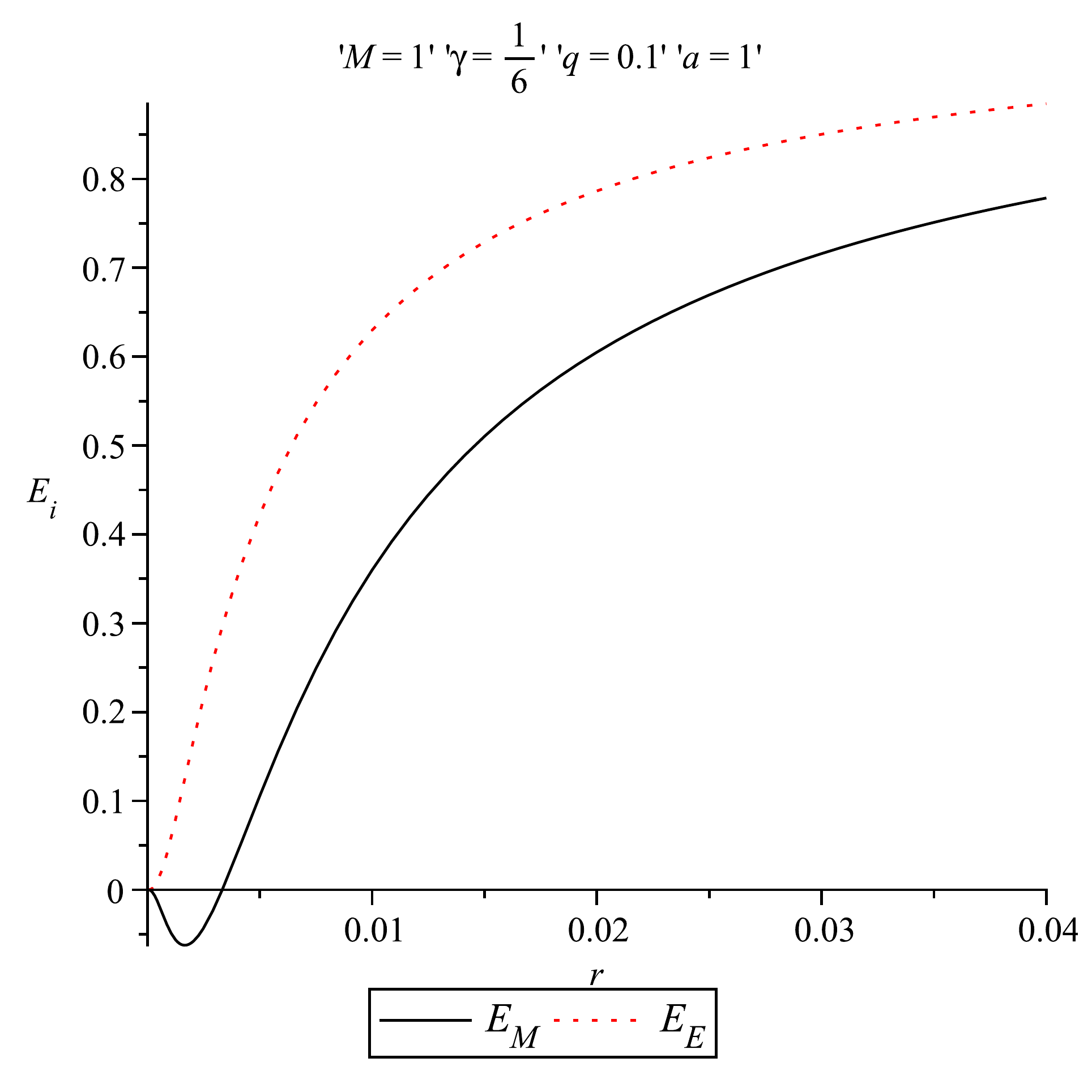}
\caption{Energy distributions in the
Einstein and M\o ller prescriptions  near the origin for $a=1$ and $\gamma =\frac{1}{6}$.}
\label{fig1}
\end{figure}

\begin{figure}[!htb]
\centering
\includegraphics[width=7cm]{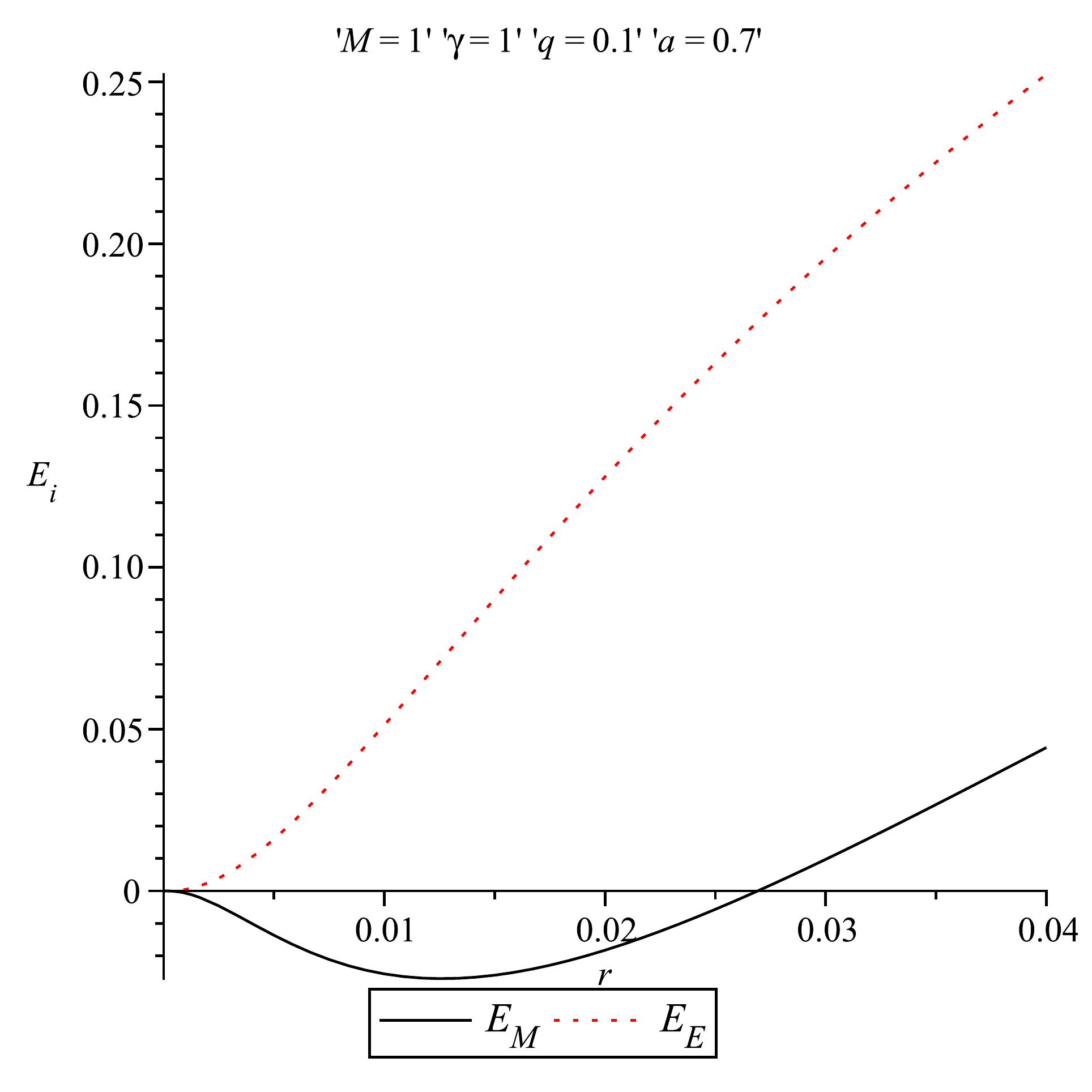}
\caption{Energy distributions in the
Einstein and M\o ller prescriptions  near the origin for $a=0.7$ and $\gamma =1$.}
\label{fig1}
\end{figure}

In Fig. 7 the energy distributions in the Einstein
prescription $E_{E}$ and in the M\o ller prescription $E_{M}$ are presented, as a function of $r$, for the
particular case $a=1$ and $\gamma =\frac{1}{6}$, while in Fig. 8 the same functions for the same values of the parameters are presented near the origin.

Looking closer at the behaviour of the energy distribution near zero, it is seen from Fig.~2 that the Einstein energy 
tends to zero from positive values, as expected from expression (27), and it is an  increasing function of $r$. The M\o ller energy, on the other hand, tends also to zero but, according to Fig.~4, close enough to zero, i.e. for $r\lessapprox 0.04$, it acquires negative values before reaching zero, a result which is supported by (30).  For values of $r$ greater than $0.04$, the M\o ller energy is positive and increasing. 

In Fig. 5 and Fig. 6, we consider the particular value $a=0.5$ as an example in order to 
compare the two energies obtained and we conclude that
the Einstein energy $E_{E}$ is everywhere greater than the M\o ller energy.
In fact, one can advocate that the positive energy region can be used for the effect of a 
convergent gravitational lens \cite{Claudel}-\cite{Virb2009}.

The negativity of the energy distribution in the case of the M\o ller prescription, for a range of values of $r$, $a$, $\gamma$, $M$ and $q$ pinpoints the difficulty of a physically meaningful interpretation of the energy in
certain regions. The values of $r$ where the energy distribution becomes negative depend on the roots of the equation $1-\frac{3\,\gamma \left( \frac{q^{2}}{Mr}
\right) ^{a}}{1+\gamma \left( \frac{q^{2}}{M\,r}\right) ^{a} }=0$ (see Eq.~(30)). In fact, the energy becomes negative for $r^a <2\gamma (\frac{q^2}{M})^a$. As an example, in Fig. 9 we see that the energy
distribution becomes negative for $0<r<0.0275$ if we choose  $a=0.7$ and $\gamma=1$. For larger values of $r$, the energy distribution becomes positive and then it increases.

As pointed out in the Introduction, a modern viewpoint regarding the energy-momentum localization is represented by a quasi-local\break energy-momentum associated with a closed 2-surface. The pseudotensors have this property, mostly since Penrose developed his definition of quasi-local mass \cite{Penrose}. Further, according to \cite{ Szabados}, since a generally accepted expression for the energy-momentum has not been found until today, some important criteria for the quasi-local energy-momentum expressions must be satisfied, basically (a) the expressions should yield the standard values at spatial infinity, and (b) they should not give negative values for the energy. In the case of the black hole solution that satisfies the weak energy condition and is under study in the present work, both the Einstein and M\o ller energy-momentum complexes do yield the standard values at infinity. In fact, in the Einstein prescription the energy takes only positive values, while in the case of the M\o ller prescription the energy takes negative values when $r$ satisfies the inequality $r^a <2\gamma (\frac{q^2}{M})^a$. We believe that this apparent weakness of the M\o ller prescription can be justified by the properties of the particular metric considered. Indeed, a similar behaviour of the M\o ller energy-momentum complex was found in \cite{Radinschi2017} and \cite{Radinschi2016}. This strange behaviour is attributed to the particularities of these black hole solutions originating in the coupling of the gravitational field to non-linear electrodynamics.
 
Although the energy distribution in the M\o ller prescription does not exhibit the desired behaviour for every value of $r$, we can conclude that the two prescriptions used in the present work still constitute instructive tools for the energy-momentum localization. Given the problem that arises from the M\o ller prescription, it is challenging to employ other pseudo-tensorial prescriptions available, as well as the  teleparallel equivalent theory of general relativity, for further investigation of the energy-momentum localization in the context of the four-dimensional spherically symmetric and charged, non-singular black hole solution satisfying the weak energy condition. We consider facing this challenge in a future work.

\end{document}